\documentclass[twocolumn,showpacs,preprintnumbers,amsmath,amssymb,pre]{revtex4-1}

\usepackage{graphicx}
\usepackage{dcolumn}
\usepackage{bm}
\usepackage{psfrag}

\begin{document}

\preprint{APS/123-QED}

\title{Magnetic energy dissipation and mean magnetic field generation in planar
convection driven dynamos}

\author{A. Tilgner}

\affiliation{Institute of Geophysics, University of G\"ottingen,
Friedrich-Hund-Platz 1, 37077 G\"ottingen, Germany }

\date{\today}

\begin{abstract}
A numerical study of dynamos in rotating convecting plane layers is presented
which focuses on magnetic energies and dissipation rates, and the generation of
mean fields (where the mean is taken over horizontal planes). The scaling of the
magnetic energy with the flux Rayleigh number is different from the scaling
proposed in spherical shells, whereas the same dependence of the magnetic
dissipation length on the magnetic Reynolds number is found for the two
geometries. Dynamos both with and without mean field exist in rapidly rotating
convecting plane layers.
\end{abstract}

\maketitle

It is generally assumed that celestial bodies without a fossil magnetic field
left over form the birth of the object create their magnetic fields through the
dynamo effect, and that in most bodies, convection is driving the motion of the
fluid conductor whose kinetic energy is transformed into magnetic energy by the
dynamo effect. Spherical shells and plane layers are two geometries in which
convection driven dynamos are conveniently studied with numerical simulations.
More attention has been paid to the spherical shell because of its greater geo-
and astrophysical relevance and a larger data base exists for this geometry. A
scaling for the magnetic field energy derived from these simulations
\cite{Christ06} has matched observations well \cite{Christ09} and has even
been invoked for mechanically driven dynamos \cite{Dwyer11,Lebars11}, which
raises the question of how universal this scaling is. An obvious test is to
compare convection dynamos in spherical shells with the most closely related
standard problem, which is convection driven dynamos in plane layers. Rotating
convection in spherical shells is inhomogeneous in the sense that the region
inside the cylinder tangent to the inner core and coaxial with the rotation axis
behaves differently from the equatorial region, and boundaries are curved.
Convection in a plane layer with its rotation axis perpendicular to the plane of
the layer may be viewed as a model for a small region surrounding the poles of a
spherical shell. Are the physics in this region representative for the rest?
This question is the motivation to look at the scaling of magnetic energy and
energy dissipation in plane layer dynamos and to compare the results with data
from simulations in spherical shells.

Field morphologies are more difficult to compare. Dynamos in spherical shells
are frequently classified according to whether they produce a magnetic field
dominated by its dipole component (in which case they are a candidate for a
model of the geodynamo) or not. A similar distinction can be made among plane
layer dynamos: They either produce a mean field, obtained by averaging over
horizontal planes, or not. Even though this issue is not obviously analogous to
the question of the dominating dipole field in the sphere, the end of this paper
will be devoted to showing that a transition in plane layer dynamos separates
dynamos generating mean fields from those who do not.

The model and the numerical method used here are the same as in ref.
\onlinecite{Tilgne12b} and are briefly reviewed here for completeness. The
parameters of the numerical runs are the same, too, except for some additional
simulations at larger aspect ratios. Consider a plane layer with boundaries
perpendicular to the $z$ axis. Rotation $\Omega$ and gravitational acceleration
$g$ are parallel and antiparallel to this axis, respectively. The fluid in the
layer has density $\rho$, kinematic viscosity $\nu$,
thermal diffusivity $\kappa$, thermal expansion coefficient $\alpha$, and
magnetic diffusivity $\lambda$. The boundaries are located in the planes $z=0$
and $z=d$ and periodic boundary conditions are applied in the lateral directions
imposing the periodicity lengths $l_x$ and $l_y$ along the $x$ and $y$
directions. In all simulations, $l_x=l_y$, and the aspect ratio $A$ is defined as
$A=l_x/d$. Four additional control parameters govern magnetic rotating
convection within the Boussinesq approximation, namely the Rayleigh number
$\mathrm{Ra}$, the Ekman number $\mathrm{Ek}$, the Prandtl number $\mathrm{Pr}$,
and the magnetic Prandtl number $\mathrm{Pm}$. They are defined by
\begin{equation}
\mathrm{Ra}=\frac{g \alpha \Delta T d^3}{\kappa \nu} ~~~,~~~
\mathrm{Ek}=\frac{\nu}{\Omega d^2} ~~~,~~~
\mathrm{Pr}=\frac{\nu}{\kappa} ~~~,~~~
\mathrm{Pm}=\frac{\nu}{\lambda}
\end{equation}
where $\Delta T$ is the temperature difference between bottom and top
boundaries. With $d$, $d^2/\kappa$, $\kappa/d$, $\rho \kappa^2/d^2$,
$\Delta T$ and $\sqrt{\mu_0 \rho} \kappa/d$ as units of length, time, velocity,
pressure, temperature difference from the temperature at $z=d$,
and magnetic field, respectively, the nondimensional equations for the velocity
field $\bm v(\bm r,t)$ as a function of position $\bm r$ and time $t$, the
magnetic field $\bm B(\bm r,t)$ and the temperature field $T(\bm r,t)$, are
given by:
\begin{equation}
\partial_t \rho + \nabla \cdot \bm v = 0
\label{eq:conti_BQ}
\end{equation}
\begin{equation}
\begin{split}
\partial_t \bm v +(\bm v \cdot \nabla) \bm v
+ 2 \frac{\mathrm{Pr}}{\mathrm{Ek}} \hat{\bm z} \times \bm v = \\
- c^2 \nabla \rho +
\mathrm{Pr} ~ \mathrm{Ra} ~\theta \hat{\bm z} + \mathrm{Pr} \nabla^2 \bm v
+ (\nabla \times \bm B) \times \bm B
\end{split}
\label{eq:NS_BQ}
\end{equation}
\begin{equation}
\partial_t \theta + \bm v \cdot \nabla \theta -v_z =\nabla^2 \theta
\label{eq:T_BQ}
\end{equation}
\begin{equation}
\partial_t \bm B +\nabla \times (\bm B \times \bm v) =
\frac{\mathrm{Pr}}{\mathrm{Pm}} \nabla^2 \bm B
\label{eq:induc_BQ}
\end{equation}
\begin{equation}
\nabla \cdot \bm B = 0
\label{eq:div_BQ}
\end{equation}
where $\theta$ is the deviation from the conductive temperature profile. The
numerical code implements an artificial compressibility method \cite{Chorin67}
with the equation of state $p=c^2 \rho$, with pressure $p$ and
sound speed $c$. The standard Boussinesq equations, with eq. (\ref{eq:conti_BQ})
replaced by $\nabla \cdot \bm v = 0$ and with the term $-c^2 \nabla \rho$ replaced
by $\nabla p$ in eq. (\ref{eq:NS_BQ}), are recovered in the limit of $c$ tending
to infinity. In all simulations, $c$ was chosen large enough to approximate well
the Boussinesq equations \cite{Tilgne12b}.

Eq. (\ref{eq:conti_BQ}) is the full continuity equation linearized around a
density equal to 1, $\rho$ being the density perturbation. Only $\nabla \rho$
enters the momentum equation so that we can set the unperturbed density to an
arbitrary constant. The system with the linearized continuity equation reduces
to the same Boussinesq limit for $c$ tending to infinity as the full system, it
also satisfies conservation of mass, and it is computationally more efficient
because it avoids round off errors in the term $1+\rho$ appearing in the full
continuity equation.

One may also wonder if it would not be more efficient to simulate the Boussinesq
equations directly. Suppose we are content to approximate the Boussinesq solutions
to an accuracy of $1\%$ because we expect errors due to limited time averaging
of larger magnitude. The error introduced by a finite sound speed is of the
order $(U/c)^2$, where $U$ is the typical flow velocity. We thus need $c \approx
10U$ for the desired accuracy. With an explicit time stepping method, the time
step will need to be 10 times smaller for the artificial compressibility method
than for the simulations of the Boussinesq equations, assuming the
advection CFL criterion limits the size of the time step. However, every time
step solving the Boussinesq equations requires the solution of a Poisson
equation. One therefore has to compare the execution time of 10 explicit time
steps and one Poisson inversion to decide which method is better suited. The
computations presented here solved eqs. (\ref{eq:conti_BQ}-\ref{eq:div_BQ}) with 
a finite difference method implemented on
graphical processing units \cite{Tilgne12b}, which are highly parallel with
relatively slow communication between some components of the board, so that the
artificial compressibility method was favored.

The boundary conditions implemented at the top and bottom boundaries were fixed
temperature ($\theta=0$), free slip ($v_z = \partial_z v_y = \partial_z v_x =
0$), and a perfect conductor was assumed outside the fluid layer ($B_z =
\partial_z B_y = \partial_z B_x =0$). 

Spatial resolution was up to $256^3$ points. In all
runs, $\mathrm{Pr}$ was set to $0.7$, and $\mathrm{Pm}$ to either 1 or 3. For
both $\mathrm{Pm}$, the $\mathrm{Ek}$ of $2 \times 10^{-4}$, $2 \times 10^{-5}$, and $2
\times 10^{-6}$ have been simulated. For each of the six combinations of
$\mathrm{Pm}$ and $\mathrm{Ek}$, $\mathrm{Ra}$ was varied from its critical
value to up to 100 times critical for $\mathrm{Ek}=2
\times 10^{-4}$ and three times critical for $\mathrm{Ek}=2
\times 10^{-6}$. The typical length scale of rotating convection varies with
$\mathrm{Ek}$ as $\mathrm{Ek}^{1/3}$ near the onset of convection and throughout
much of the range of Rayleigh numbers investigated here \cite{Schmit10}.
Accordingly, the aspect ratio $A$ was chosen to be $A=1$, $1/2$
and $1/4$ for $\mathrm{Ek}=2 \times 10^{-4}$, $2 \times 10^{-5}$ and 
$2 \times 10^{-6}$, respectively. The aspect ratio dependence of the mean
magnetic field will be discussed towards the end of the paper.

The densities of kinetic and magnetic energies, $e_{\mathrm{kin}}$ and $e_B$, are given by
\begin{equation}
e_{\mathrm{kin}}= \frac{1}{V} \int \frac{1}{2} \bm v^2 dV ~~~,~~~
e_B= \frac{1}{V} \int \frac{1}{2} \bm B^2 dV,
\end{equation}
where the integration extends over the entire fluid volume $V$. If we denote the
time average by angular brackets, one can compute average energy densities
$E_{\mathrm{kin}}$ and $E_B$ from $E_{\mathrm{kin}}=\langle e_{\mathrm{kin}}
\rangle$ and $E_B=\langle e_B \rangle$ as well as the Reynolds number 
$\mathrm{Re}$ and the magnetic Reynolds number $\mathrm{Rm}$ from 
\begin{equation}
\mathrm{Re}=\langle \sqrt{2 e_{\mathrm{kin}}} \rangle / \mathrm{Pr}
~~~,~~~
\mathrm{Rm}=\mathrm{Re} \, \mathrm{Pm}.
\end{equation}

In the previous study of this model \cite{Tilgne12b}, it was found that there is
a transition at $\mathrm{Rm}\, \mathrm{Ek}^{1/3}=13.5$. The combination
$\mathrm{Rm}\, \mathrm{Ek}^{1/3}$ is proportional to the magnetic Reynolds number based on the
size of a columnar vortex near the onset of convection. As the Rayleigh number
is increased starting from small values, the growth rate of kinematic dynamos first
increases, then goes through a minimum at $\mathrm{Rm}\, \mathrm{Ek}^{1/3}=13.5$
and then increases again. The growth rate is not a monotonic function of neither
$\mathrm{Rm}$ nor $\mathrm{Ra}$ at constant $\mathrm{Ek}$.
The amplitude of the saturated magnetic field obeys
different scaling laws below and above this transition. These are given in ref.
\onlinecite{Tilgne12b} in terms of $\mathrm{Rm}$, $\mathrm{Ek}$,
and $\mathrm{Pm}$. The $\mathrm{Rm}$ is not a control parameter
of the problem, but it is more accessible to observations than $\mathrm{Ra}$, so
that these scaling laws are of interest even if they are not expressed in terms
of control parameters only.

Another parameter of greater relevance to observations than $\mathrm{Ra}$ is the
flux Rayleigh number, $\mathrm{Ra_f}$, based on the heat flux. If the fluid is
at rest, the heat flux across the layer is purely diffusive and given by 
$\kappa \rho c_p \Delta T/d$ where $c_p$ is the heat capacity at constant
pressure. When convection sets in, the heat flux may be written as
$\kappa \rho c_p \Delta T/d + Q_{\mathrm{adv}}$, where $Q_{\mathrm{adv}}$ is the
difference between the actual heat flux and the diffusive heat flux through the
fluid at rest. The Nusselt number $\mathrm{Nu}$ is defined as
\begin{equation}
\mathrm{Nu}=1+Q_{\mathrm{adv}}/(\kappa \rho c_p \Delta T/d) 
\end{equation}
and
\begin{equation}
\mathrm{Ra_f} = \mathrm{Ra}\, \mathrm{(Nu-1)}\, \mathrm{Ek^3}/\mathrm{Pr} =
(g \alpha Q_{\mathrm{adv}})/(\rho c_p \Omega^3 d^2).
\end{equation}
The flux Rayleigh number is
independent of diffusivities, and the heat flux is better
constrained by observations than the temperature difference $\Delta T$.

\begin{figure}
\includegraphics[width=8cm]{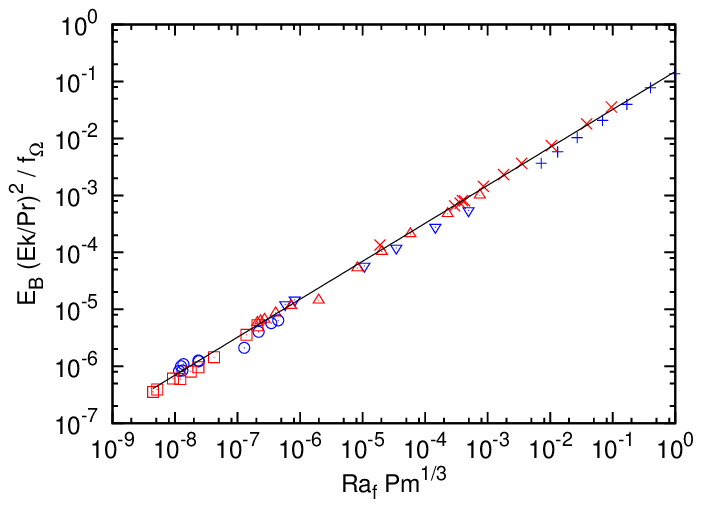}
\includegraphics[width=8cm]{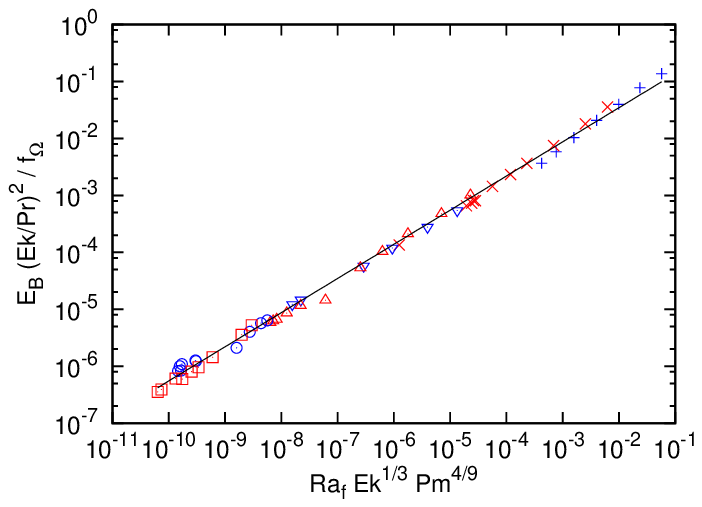}
\caption{(Color online)
$(E_B/f_\Omega) (\mathrm{Ek}/\mathrm{Pr} )^2$
as a function of $\mathrm{Ra_f} \, \mathrm{Pm}^{1/3}$ (top panel) and
$\mathrm{Ra_f} \, \mathrm{Ek}^{1/3} \, \mathrm{Pm}^{4/9}$
(bottom panel).
Results for $Pm=1$ are shown
in blue and those for $Pm=3$ are in red. For $Pm=1$, the Ekman numbers of
$2 \times 10^{-4}$, $2 \times 10^{-5}$, and $2 \times 10^{-6}$ are indicated by
the plus sign, triangle down, and circle, respectively, whereas for $Pm=3$, the
same Ekman numbers are indicated by the x sign, triangle up, and square. The
straight lines show power laws with the exponents 2/3 (top panel) and 3/5
(bottom panel).}
\label{fig:EB_Raf_1}
\end{figure}

\begin{figure}
\includegraphics[width=8cm]{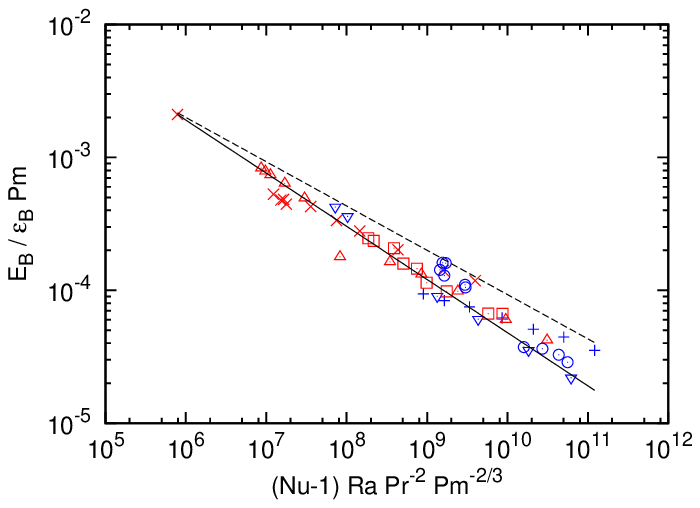}
\includegraphics[width=8cm]{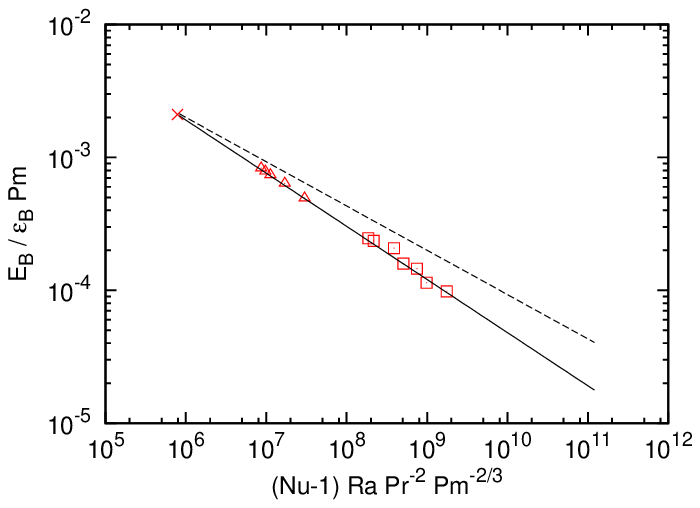}
\caption{(Color online)
$(E_B/\epsilon_B)\mathrm{Pm}$ as a function of 
$\mathrm{Ra} \, (\mathrm{Nu}-1) \, \mathrm{Pr}^{-2} \, \mathrm{Pm}^{-2/3}$ with
the same symbols as in fig. \ref{fig:EB_Raf_1}. The straight lines indicate
power laws with the exponents $-2/5$ (solid line) and $-1/3$ (dashed line). The
bottom panel contains the same data as the top panel but shows only points below
the transition with $\mathrm{Rm}\, \mathrm{Ek}^{1/3}<13.5$ and $\mathrm{Pm}=3$.}
\label{fig:lB_Nu}
\end{figure}

It would be interesting to know a relation between the saturation magnetic field
strength and $\mathrm{Ra_f}$. From their simulations in spherical shells,
Christensen and Aubert \cite{Christ06} find 
$(E_B/f_\Omega) (\mathrm{Ek}/\mathrm{Pr})^2 = (0.76 \, \mathrm{Ra_f}^{0.32}
\mathrm{Pm}^{0.11})^2$
where $f_\Omega$ is the ratio of ohmic to total dissipation, which in the units
used here is given by
\begin{equation}
f_\Omega = \frac{\epsilon_B/\mathrm{Pm}}{(\epsilon_v + \epsilon_B /\mathrm{Pm})}
\end{equation}
with
\begin{equation}
\epsilon_B = \frac{1}{V} \int < (\nabla \times \bm B)^2>  dV
\end{equation}
and
\begin{equation}
\epsilon_v = \frac{1}{V} \int < (\partial_j v_i) (\partial_j v_i) > dV,
\end{equation}
where summation over repeated indices is implied and the integration extends
over the whole computational volume. The form of this scaling comes from an
attempt to determine the magnetic field strength not from a balance of forces
but from energy considerations. One can derive from the equations of evolution
(\ref{eq:conti_BQ}-\ref{eq:div_BQ}) (in the limit of large sound speed $c$, i.e.
in the standard Boussinesq limit) the energy budget
\begin{equation}
\epsilon_v + \frac{\epsilon_B}{\mathrm{Pm}} = (\mathrm{Nu}-1) \mathrm{Ra}.
\label{eq:energy_budget}
\end{equation}
For the spherical dynamo models with the radial variation of gravity usually
simulated, an exact equation of the same structure is not available, but a fit in
ref. \cite{Christ06} shows that the total dissipation is still approximately
proportional to $(\mathrm{Nu}-1) \mathrm{Ra}$. The purely ohmic dissipation is
related to the total dissipation by the factor $f_\Omega$ by definition, and
$E_B/\epsilon_B$ is the square of a magnetic length
scale, $l_B$, with 
\begin{equation}
l_B = \sqrt{E_B/\epsilon_B}. 
\label{eq:def_lB}
\end{equation}
The magnetic dissipation time,
defined as the ratio of magnetic energy and ohmic dissipation, made
nondimensional with the ohmic diffusion time, is also given by $E_B/\epsilon_B$.
Ref. \cite{Christ06} finds an acceptable fit for $l_B$ as a function of the
control parameters of the flow, which together with the fit for the total
dissipation rate as a function of $(\mathrm{Nu}-1) \mathrm{Ra}$ leads to a
relation between $E_B/\epsilon_B$ and the control parameters. 
A more elaborate fitting procedure \cite{Stelze13} in which one searches directly
a power law fit for $E_B/f_\Omega$ as a function of $\mathrm{Ra_f}$,
$\mathrm{Ek}$ and the Prandtl numbers leads to
$(E_B/f_\Omega) (\mathrm{Ek}/\mathrm{Pr})^2 = (0.60 \, \mathrm{Ra_f}^{0.31}
\mathrm{Pm}^{0.16})^2$. For the purpose of the discussion below, we can round
the exponents to
\begin{equation}
\frac{E_B}{f_\Omega} \left( \frac{\mathrm{Ek}}{\mathrm{Pr}} \right)^2
\propto \left( \mathrm{Ra_f} \, \mathrm{Pm}^{1/3} \right) ^{2/3}.
\label{eq:EB_Raf_1}
\end{equation}
The data available for the plane layer will not allow us to determine an
exponent
for $\mathrm{Pm}$, and the analysis of the $\mathrm Ra_f$ dependence will not
depend on discrepancies of 0.01 in the exponent. Note also that the factor
$\mathrm{Ek}/\mathrm{Pr}$ on the left hand side is due to the different units of
magnetic field used here and in ref. \onlinecite{Christ06}.
Eq. (\ref{eq:EB_Raf_1}) has no predictive power for $E_B$ unless one guesses
$f_\Omega$. However, an upper bound for $f_\Omega$ is 1, resulting in an upper
bound for $E_B$ if $f_\Omega$ is set to 1 in Eq. (\ref{eq:EB_Raf_1}).

Figure \ref{fig:EB_Raf_1} shows $(E_B/f_\Omega) (\mathrm{Ek}/\mathrm{Pr})^2$ as
a function of $\mathrm{Ra_f} \, \mathrm{Pm}^{1/3}$ for the plane layer dynamos
and eq. (\ref{eq:EB_Raf_1}) seems to provide a satisfying fit. Remarkably, there
is no trace of a transition between different types of dynamos in this plot.
However, one can simplify Eq. (\ref{eq:EB_Raf_1}) by using the energy budget
(\ref{eq:energy_budget}) in order to obtain
\begin{equation}
\frac{E_B}{\epsilon_B} \mathrm{Pm} \propto
\left( \mathrm{Ra} \, (\mathrm{Nu}-1) \, \mathrm{Pr}^{-2} \, \mathrm{Pm}^{-2/3}
\right) ^{-\alpha}.
\label{eq:lB_Nu}
\end{equation}
with $\alpha=1/3$. This equation is simpler than eq. (\ref{eq:EB_Raf_1}) because
common factors $\mathrm{Ra} (\mathrm{Nu}-1)$ and $\mathrm{Ek}$ are removed.
Figure \ref{fig:lB_Nu} shows $(E_B/\epsilon_B)\mathrm{Pm}$ as a function of
$\mathrm{Ra}(\mathrm{Nu}-1)\mathrm{Pr}^{-2}\mathrm{Pm}^{-2/3}$. Because of the
removal of the common factors, the data points spread over fewer decades and it
becomes apparent that $\alpha=1/3$ is not an acceptable exponent, neither as a
fit to the data cloud as a whole, nor to the points below the transition at
$\mathrm{Rm}\, \mathrm{Ek}^{1/3}=13.5$, nor to individual series of simulations
at $\mathrm{Ek}$, $\mathrm{Pm}$ and $\mathrm{Pr}$ constant. Instead, the best
fitting exponent  is close to $\alpha=2/5$. This exponent describes the
dependence on
$\mathrm{Ra}(\mathrm{Nu}-1)$. The dependence on $\mathrm{Pm}$ and $\mathrm{Pr}$
is not seriously tested by the data.

We can now reinflate eq. (\ref{eq:lB_Nu}) for $\alpha=2/5$
with the help of the energy budget to
obtain a relation analogous to eq. (\ref{eq:EB_Raf_1}), which becomes
\begin{equation}
\frac{E_B}{f_\Omega} \left( \frac{\mathrm{Ek}}{\mathrm{Pr}} \right)^2
= 0.55 \left( \mathrm{Ra_f} \, \mathrm{Ek}^{1/3} \, \mathrm{Pm}^{4/9} \right)
^{3/5}.
\label{eq:EB_Raf_2}
\end{equation}
where the prefactor is taken from fig. \ref{fig:EB_Raf_1} which shows eq.
(\ref{eq:EB_Raf_2}) to be a satisfactory fit, again. An $\mathrm{Ek}$ dependence
of the right hand side therefore appears in eq. (\ref{eq:EB_Raf_2}). In the
spherical models on the contrary, the best fit does not contain any
$\mathrm{Ek}$ dependence (see table 4 of ref. \onlinecite{Stelze13}).

\begin{figure}
\includegraphics[width=8cm]{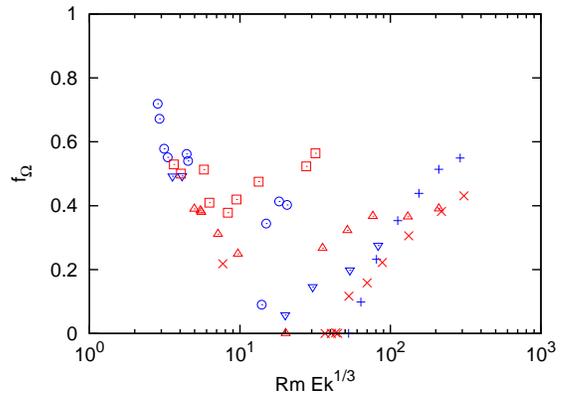}
\caption{(Color online)
$f_\Omega$ as a function of $\mathrm{Rm} \, \mathrm{Ek}^{1/3}$ with
the same symbols as in fig. \ref{fig:EB_Raf_1}.}
\label{fig:f_Ohm}
\end{figure}

For completeness, fig. \ref{fig:f_Ohm} plots $f_\Omega$ as a function of
$\mathrm{Rm} \, \mathrm{Ek}^{1/3}$. It is plausible that $f_\Omega$ is small for
dynamos close to the onset in the case of a supercritical bifurcation and that
$f_\Omega$ approaches 1 as the magnetic field grows stronger. However,
$f_\Omega$ is already 0.7 at the smallest $\mathrm{Rm} \,
\mathrm{Ek}^{1/3}$ in fig. \ref{fig:f_Ohm}. This supports the scenario of a
subcritical convection driven dynamo in plane layers \cite{Stellm04, Jones00}.
According to \cite{Tilgne12b}, a second type of dynamo operates for $\mathrm{Rm}
\, \mathrm{Ek}^{1/3} > 13.5$, and in this range, $f_\Omega$ is increasing as a
function of $\mathrm{Rm} \, \mathrm{Ek}^{1/3}$ as expected.
When scalings of $E_B$ are
sought in terms of $\mathrm{Rm}$ and $\mathrm{Ek}$, these two types of dynamos
have to be considered separately \cite{Tilgne12b}, but they can be fitted
simultaneously in a graph of $E_B/f_\Omega$ like fig. \ref{fig:EB_Raf_1} because
the complications of the transition are hidden in $f_\Omega$.

\begin{figure}
\includegraphics[width=8cm]{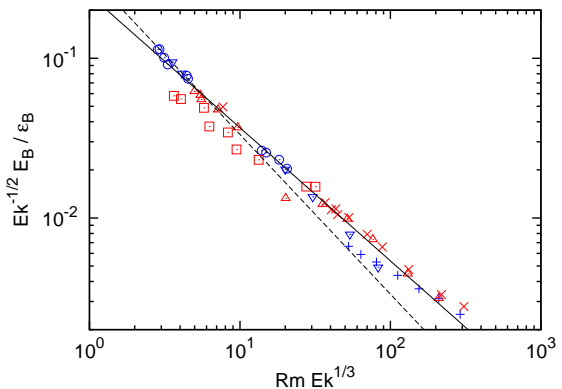}
\caption{(Color online)
$\mathrm{Ek}^{-1/2} E_B/\epsilon_B$ as a function of 
$\mathrm{Rm} \, \mathrm{Ek}^{1/3}$ with
the same symbols as in fig. \ref{fig:EB_Raf_1}.  The straight lines indicate
power laws with the exponents $-5/6$ (solid line) and $-1$ (dashed line).
}
\label{fig:lB_Rm}
\end{figure}

\begin{figure}
\includegraphics[width=8cm]{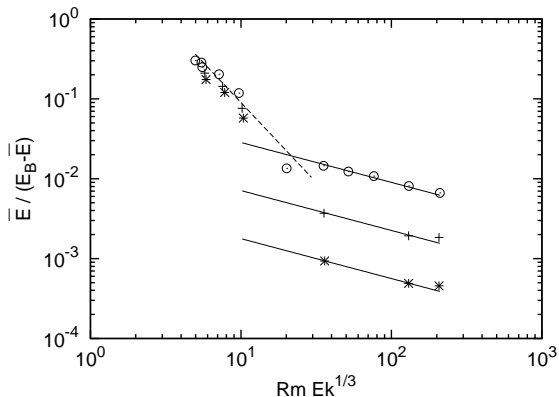}
\caption{
$\bar E / (E_B-\bar E)$ as a function of $\mathrm{Rm} \, \mathrm{Ek}^{1/3}$ for
$A=0.5$ (circles), 1 (crosses) and 2 (stars). The dashed line follows the
prediction of first order smoothing and shows $\bar E / (E_B-\bar E) \propto
(\mathrm{Rm} \, \mathrm{Ek}^{1/3})^{-2}$. The solid lines plot the functions
$0.09/\sqrt{x}$, $0.09/(4\sqrt{x})$, and $0.09/(16\sqrt{x})$.}
\label{fig:AR}
\end{figure}

The magnetic length scale $l_B$ introduced in eq. (\ref{eq:def_lB})
is connected to the total dissipation $\mathrm{Ra} \, (\mathrm{Nu}-1)$ in Eq.
(\ref{eq:lB_Nu}). It is more natural to seek a
relation between $l_B$ and $\mathrm{Rm}$. From spherical shell simulations, ref.
\onlinecite{Christ04} infers $E_B/\epsilon_B \propto 1/\mathrm{Rm}$, whereas a
more extended analysis \cite{Christ10} yielded $E_B/\epsilon_B \propto
\mathrm{Rm}^{-5/6} (\mathrm{Ek}/\mathrm{Pm})^{1/6}$. There is no theoretical
basis for this relationship, it is at present a purely empirical finding. 
Fig. \ref{fig:lB_Rm} shows
$\mathrm{Ek}^{-1/2} E_B/\epsilon_B$ as a function of $\mathrm{Rm} \, \mathrm{Ek}^{1/3}$ and
confirms the dependence of $l_B$ on $\mathrm{Rm}$ in $\mathrm{Rm}^{-5/6}$ which
is therefore identical in spherical and in planar geometry, and also confirms
the $\mathrm{Ek}$ dependence found in ref. \cite{Christ10} to within a factor 
$\mathrm{Ek}^{1/18}$, which is too small to be discerned in the data. 

Fig. \ref{fig:lB_Rm} shows that the behavior of the magnetic dissipation
length $l_B$ is not affected by the transition at $\mathrm{Rm} \,
\mathrm{Ek}^{1/3}=13.5$ and that it behaves the same for the two types of
dynamos, above and below the transition. 
The variable $\mathrm{Rm} \, \mathrm{Ek}^{1/3}$ does on the other hand decide on
whether a mean field is generated. The energy in the mean field, $\bar E$, is
computed as 
\begin{equation}
\bar E = \frac{1}{2} \frac{1}{V} < \int dz \left( \int dy \int dx \, \bm B
\right)^2 >.
\label{eq:def_E_mean}
\end{equation}
It is well known that close to the onset of dynamo action in rapidly rotating
plane layer convection, the generated magnetic field is dominated by its mean
field component \cite{Stellm04}. The dynamo is then accessible to the tools of
mean field magnetohydrodynamics and first order smoothing \cite{Soward74} which
predict $\bar E / (E_B-\bar E) \propto (\mathrm{Rm} \, \mathrm{Ek}^{1/3})^{-2}$. 
In the simulations presented here, the ratio $\bar E / (E_B-\bar E)$ was smaller
than 0.01 at the highest $\mathrm{Rm} \, \mathrm{Ek}^{1/3}$.
The simulations at
$\mathrm{Ek}=2 \times 10^{-5}$ and $\mathrm{Pm}=3$ have been complemented by
simulations at different aspect ratios. Most points have been obtained at an
aspect ratio of 0.5, and a few points have been added for aspect ratios 1 and 2.
The result is shown in fig. \ref{fig:AR}. If the aspect ratio is increased for 
points below the transition at $\mathrm{Rm} \, \mathrm{Ek}^{1/3}<13.5$, one
observes variations in both $\mathrm{Rm}$ and $\bar E$ which increase as one
approaches the transition. However, the variation in $\bar E / (E_B-\bar E)$ is
always less than by a factor of 2 even if the aspect ratio changes by a factor
of 4. Above the transition, on the other hand, an increase of the aspect ratio
$A$ by a factor of 2 always reduces $\bar E / (E_B-\bar E)$ by a factor of 4.
This behavior is readily understood if one assumes that these dynamos do not genuinely
generate a mean field, but that the statistical fluctuations of the local field
do not cancel exactly in a volume of finite size. Assume that the magnetic field
has a correlation length $l_c$. The number of independent degrees of freedom in
a plane of cross section $A \times A$ is $(A/l_c)^2$. The mean field computed in
each plane is the sum of $(A/l_c)^2$ random numbers drawn from a probability
distribution with a width proportional to $\sqrt{E_B}$, so that
$\bar E / (E_B-\bar E) \approx \bar E / E_B \propto (l_c/A)^2$.
Doubling $A$ thus reduces $\bar E / (E_B-\bar E)$ by a factor of 4. 


The evidence thus points at a dynamo without a mean magnetic field above the
transition, even though in any numerical realization, the mean field is not
exactly zero but depends on the aspect ratio. Below the transition, the dynamo
does generate a mean field, but as its amplitude is small, the contribution from
the statistical fluctuations of the mean field introduces some aspect ratio
dependence in these dynamos as well.

Favier and Bushby \cite{Favier13} also found in their simulations of
dynamos in rotating compressible convection a mean field which decreases with
increasing aspect ratio, so that the mean field detected in these simulations
may well be a statistical feature as described above. Cattaneo and Hughes
\cite{Cattan06} simulate dynamos which produce magnetic energy spectra which
peak at small scales suggesting a dynamo process at small scales (similarly
to \cite{Favier13}). They for
example present a case with $\mathrm{Rm} \, \mathrm{Ek}^{1/3}$ around 200 (which
is clearly above the transition) at the relatively large $\mathrm{Ek}$ of $2.8
\times 10^{-3}$ and an aspect ratio of 10 and find as expected a small value for
$E_B/\epsilon_B$ on the order of $10^{-4}$.

Large mean fields were observed on the other hand in refs. \cite{Stellm04,
Jones00, Rotvig02}. Stellmach and Hansen \cite{Stellm04} used the exact same
model as here, but simulated Rayleigh numbers closer to onset than in the
present study, so that the existence of an important mean field is not
surprising. Refs. \cite{Jones00, Rotvig02} used Rayleigh numbers a few times and
up to ten times critical, and Ekman numbers comparable to these in the present
study, so that these dynamos should be examples of dynamos below the transition.
The authors found mean fields about 2-3 times as large as here. One may
speculate that this is due to different boundary conditions: For the perfectly
conducting boundaries used here, the average over $z$ of the mean field must be
zero \cite{Jones00}, but this constraint does not exist for the insulating boundaries used in
refs. \cite{Jones00, Rotvig02}.

In summary,
convection in rotating plane layers supports dynamos both with and without a mean
field. 
The scaling exponents for the energy and the magnetic dissipation length inferred
from simulations in spherical shells at first glance fit perfectly well the data
from the plane layer. However, closer inspection reveals the field energy
scaling proposed for the spherical shell to be unacceptable for the plane layer data.
Of course, more aspects of the model than the boundary geometry have been
changed in going from the usual spherical dynamo simulation to the plane layer
model presented here, such as the spatial variation of gravity and the magnetic
boundary conditions, and it is not yet possible to tell which of those features is
relevant for the magnetic field scalings. The present work at any rate leads us 
to also expect differences
between different spherical models, such as models with
different ratios of outer and inner radii, with different radial dependencies
of gravity, or with different boundary conditions.


%

\end{document}